# Stability Analysis of Relativistic Polytropes

A. S. Saad[1, 2], M. I. Nouh[2], A. A. Shaker[2] and T. M. Kamel[2]

[1]Department of Mathematics, Deanship of Educational Services, Qassim University, Buraidah, Saudi Arabia
Email: saad6511@gmail.com
[2]Department of Astronomy, National Research Institute of Astronomy and Geophysics, 11421 Helwan, Cairo, Egypt
Email: abdo_nouh@hotmail.com

**Abstract:**

A main question in astrophysics and cosmology has been the severe stability of the astrophysical objects, whether a particular equilibrium configuration is stable. In this article, we study the relativistic self-gravitating, hydrostatic spheres with a polytropic equation of state $P = K\rho^{1+1/n}$, considering structures with the polytropic indices n=1, 1.5, 2, 2.5, 3 and illustrates the results for the relativistic parameters $\sigma$=0 (0.01) 0.75. We determined the critical relativistic parameter at which the mass of the polytrope has a maximum value and represents the first mode of radial instability. For n=1 (0.5)-2.5, stable relativistic polytropes occur for $\sigma$ less than the critical values 0.42, 0.20, 0.10, and 0.04 respectively, while unstable relativistic polytropes are obtained when the relativistic parameter $\sigma$ is greater than the same values. When n=3.0 and $\sigma > 0.5$, energetically unstable solutions have occurred. The results of critical values obtained in this paper for different polytropic indices are in full agreement with those evaluated by several authors. Comparisons between analytical and numerical solutions of the given relativistic functions provide a maximum relative error of order $10^{-3}$.

## 1. Introduction

The polytropic models could be considered simple models of stellar structure, and it seems like we have in place all of the equations that we would need to be able to make more sophisticated stellar models by solving the equations of stellar structure. However, before one dives forward, there is a need to ask a vital question, whether the calculated models are spherically symmetric equilibria (stability of the models).

In this category, several authors have investigated the stability of the polytropic models. Bonnor (1958) finds that self-gravitating, polytropic spheres with n=3 are inconditionally stable to radial perturbations. For the first time, Chandrasekhar (1964) stated the radial stability equation. Earlier methods used to examine the stability of polytropic stars are listed in Bardeen et al. (1966). More recently, the stability of polytropes with different polytropic indices has been described by Horedt (2013) and Raga et al. (2020).

In various stellar structures as white dwarfs, neutron stars, black holes and supermasive stars and cluster of stars, relativistic effects play a significant role, Sen and Roy (1954), Sharma, (1988). Tooper (1964) has performed a relativistic analysis of the polytropic equation of state. Tooper derived the non-relativistic Lane-Emden equation from two nonlinear differential equations (Tolman-Oppenheimer, TOV). The problem of the stability of relativistic stars has a longstanding investigation through the literature, for example, Zeldovich and Novikov (1978), Shapiro and Teukolsky (1984), Takatsuka and Tamagaki (1993), Casalbuoni and Nardulli (2004), Khalilov (2002), Isayev (2015), Chu et al. (2015).

In the present paper, we examine the stability of the relativistic polytrope for different polytropic indices. An analytical solution to the TOV equation is introduced which provided us with the physical parameters of the relativistic polytrope. We investigated the critical values of the relativistic parameter that the onset of the radial instability occurs. The structure of the paper is as follows: section 2 is devoted to the formulation of the TOV equation. In section 3 we give a brief description of the analytical method used to solve the TOV equation. Section 4 deals with the obtained results. The conclusion is outlined in section 5.

## 2. The Equation of Hydrostatic Equilibrium

The interior of the symmetric star can be described in a spherical coordinate system $(r, \vartheta, \varphi)$ by the standard form of the metric (Tolman 1939, Landu & Lifshitz 1962)

$$ds^2 = e^{\nu}c^2 dt^2 - e^{\lambda}dr^2 - r^2 d\vartheta^2 - r^2 \sin^2\vartheta d\varphi^2, \tag{1}$$

where $\nu$ and $\lambda$ are functions of radius $r$. As for a fluid star, the components of the energy momentum tensor corresponding to the above metric are given by

$$T_0^0 = \rho c^2 e^{\nu}, \quad T_1^1 = P e^{\lambda}, \quad T_2^2 = P r^2, \quad T_3^3 = P r^2 \sin^2\vartheta, \tag{2}$$

where $\rho$, $P$ and $c$ are the mass density, pressure, and speed of light respectively. The time-independent gravitational equations for the line element (1) and the energy momentum tensor are

$$e^{-\lambda}\left(\frac{1}{r}\frac{d\nu}{dr} + \frac{1}{r^2}\right) - \frac{1}{r^2} = \frac{8\pi G}{c^4} P, \tag{3}$$

$$e^{-\lambda}\left(\frac{1}{r}\frac{d\lambda}{dr} - \frac{1}{r^2}\right) + \frac{1}{r^2} = \frac{8\pi G}{c^4} \rho c^2, \tag{4}$$

$$\frac{dP}{dr} = -\frac{1}{2}(P + \rho c^2)\frac{d\nu}{dr}, \tag{5}$$

where $G = 6.67 \times 10^{-8} g^{-1} cm^3 s^{-2}$ is the Newtonian gravitational constant. Equations (3), (4), and (5) together with the equation of state $\rho = \rho(P)$ represent the hydrostatic equilibrium for an isotropic general relativistic fluid sphere and can be solved to get $\lambda, \nu, P$ and $\rho$ as functions of $r$. For hydrostatic equilibrium stars, Tolman-Oppenheimer-Volkoff (TOV) general relativity equation obtained by solving Einstein's field equations has the form

$$\frac{dP}{dr} = -\frac{G\ \varepsilon(r)\ m(r)}{c^2\ r^2}\left[1 + \frac{P(r)}{\varepsilon(r)}\right]\left[1 + \frac{4\pi r^3\ P(r)}{m(r)\ c^2}\right]\left[1 - \frac{2G\ m(r)}{c^2\ r}\right]^{-1}, \tag{6}$$

where

$$m(r) = \int_0^r 4\pi \rho(r) r^2 dr,$$

is the gravitational mass interior to radius $r$ and $\varepsilon(r)$ is the internal energy density. Equation (6) is an extension of the Newtonian formalism with relativistic correction. The equation of state for a polytropic star is $P = K\rho^{1+\frac{1}{n}}$, where $n$ is the polytropic index. Tooper (1964) has shown that the TOV equation together with the mass conservation equation has the form

$$\xi^2 \frac{d\theta}{d\xi} \frac{1-2\sigma\,(n+1)\,\upsilon/\xi}{1+\sigma\,\theta} + \upsilon + \sigma\,\xi\,\theta\frac{d\upsilon}{d\xi} = 0, \tag{7}$$

$$\text{and}\quad \frac{d\upsilon}{d\xi} = \xi^2 \theta^n, \tag{8}$$

with the initial conditions

$$\theta(0)=1, \quad \upsilon(0)=0, \tag{9}$$

where

$$\theta = \rho/\rho_c, \quad \xi = rA, \quad \upsilon = \frac{A^3 m(r)}{4\pi\rho_c},$$

$$A = \left(\frac{4\pi G\rho_c}{\sigma(n+1)c^2}\right)^{1/2}, \quad \sigma = \frac{P_c}{\rho_c\,c^2} = \frac{K\rho_c^{1/n}}{c^2}, \tag{10}$$

$\sigma$ is the relativistic parameter can be related to the sound velocity in the fluid, that is because the sound velocity is given by $\upsilon_s^2 = \frac{dP}{d\rho}$ in an adiabatic expression. In equations(10) $\theta$, $\xi$ and $\upsilon$ are dimensionless parameters, while $A$ is a constant.

If the pressure is much smaller than the energy density at the center of a star (i.e. $\sigma$ tends to zero), then equation (7) reduces to

$$\xi^2 \frac{d\theta}{d\xi} + \upsilon = 0. \tag{11}$$

Equation (8) together with equation (11) reproduces the well-known Lane-Emden equation for Newtonian polytropic stars

$$\frac{1}{\xi^2}\frac{d}{d\xi}\left(\xi^2 \frac{d\theta}{d\xi}\right) + \theta^n = 0. \tag{12}$$

When $n$ tends to zero, we get the case of incompressible matter, for which the analytic solutions are possible in both relativistic and nonrelativistic cases. The nonrelativistic Lane-Emden equation has an analytical solution in a closed form for only $n = 0$, 1 and 5, however, this is not possible for relativistic equation, and recourse must be had to numerical integration (Tooper 1964, Bludman 1973, Ferrari et al. 2007).

**3. Analytical Solutions of the relativistic Equations**

Nouh (2004), Nouh and Saad (2013) introduced a new analytical solution of Equations (7-8) and applying the Euler-Abel transformation (Demodovich & Maron, 1973) and then Pade approximation to the Euler-Abel transformed series (Appendix B) to accelerate the convergence of the power series solutions.

In this paper, we analyze the gravitational stability of polytropic fluid spheres based on the analytical solution of TOV equations that have already been given by Nouh and Saad (2013). We considered the cases of the polytropic index $n = 3.0, 2.5, 2.0, 1.5$ and $1.0$ for $\sigma < n/(n+1)$.

The analytical solution has the form:

$$\theta(\xi) = 1 + \sum_{k=1}^{\infty} a_k \xi^{2k}, \tag{13}$$

where

$$a_{k+1} = \frac{\sigma}{2(k+1)}(2(n+1)\gamma_{k-1} - \eta_k - \beta_k + \sigma \zeta_k) - \frac{\alpha_k}{2(k+1)(2k+3)}, \ k \geq 1,$$

$$\gamma_{k-1} = \sum_{i=0}^{k-1} f_i g_{k-i-1}, \ \eta_k = \sum_{i=0}^{k} a_i g_{k-i}, \ \beta_k = \sum_{i=0}^{k} a_i \alpha_{k-i}, \ \zeta_k = \sum_{i=0}^{k} a_i \beta_{k-i},$$

$$f_i = 2(i+1) a_{i+1}, g_i = \frac{\alpha_i}{(2i+3)}, \ \gamma_k = \sum_{i=0}^{k} f_i g_{k-i}, \tag{14}$$

$$\alpha_k = \frac{1}{k\, a_0} \sum_{i=1}^{k} (n\, i - k + i)\, a_i\, \alpha_{k-i}, \ k \geq 1, \alpha_0 = a_0^n, \text{ and } a_0 = 1.$$

From equations (10), for some values of $n$, $\sigma$ and $\rho_c$ we can determine $K$, and obtain the radius $R$ and the mass $M(R)$ from

$$R = A^{-1}\xi_1 = \left[\frac{c^2}{4\pi G}(n+1)\sigma^{(1-n)}\left(\frac{K}{c^2}\right)^n\right]^{1/2} \xi_1, \tag{15}$$

$$M = \frac{4\pi \rho_c}{A^3} v(\xi_1) = \left[\frac{1}{4\pi}\left(\frac{(n+1)c^2}{G}\right)^3 \left(\frac{K}{c^2}\right)^n\right]^{1/2} \tilde{M}, \tag{16}$$

$$\tilde{M} \equiv \sigma^{(3-n)/2}\, v(\xi_1), \tag{17}$$

$\xi_1$ is the first zero of the Lane-Emden function $\theta(\xi)$. Equation (8), can be written in the form

$$\upsilon(\xi_1) = \sum_{k=0}^{\infty} \frac{\alpha_k}{(2k+3)} \xi_1^{2k+3} \ . \tag{18}$$

The power series solution (13) is converged rapidly for the polytropic index $n \leq 2$ and the order of error between analytical and numerical solutions was of order $10^{-4}$. For $n > 2$, the series solution has utilized gives slow convergence, and calculation of the stellar mass from equations (16) and (17) indicates an increase in error. The physical range for a convergent power series can be extended with a change of the independent variable. Transformations by changing of the independent variable are utilized to improve and accelerate the series convergence in equation (18) for $n > 2$ (Pasucal 1977, Saad 2004):

$$x = 6 * \left\{ \left( 1 + \frac{1}{3} \xi^2 \right) \right\}^{1/2} \ . \tag{19}$$

## 4. Results

The results evaluated by the usage of Equations in Section 3 are utilized here to analyze the stability of relativistic polytropes for various values of the general relativity $\sigma$ and polytropic index *n*. The numerical results obtained are tabulated in Appendix A for n=1(0.5) to 3.0 and a range value of $\sigma$. Comparisons of the analytical solutions of *v*(ξ1) and $\tilde{M}(\sigma)$ to the numerical method are given in Tables A1 to A5. While Table A6 shows the critical values of $\tilde{M}(\sigma)$ due to relativistic effects for different polytropic indices.

In figures 1 to 5 we plot $\tilde{M}$ (Equation 17) as a function of the index *n* and the relativistic effect $\sigma$. The figures show an increase of $\tilde{M}$ (consequently increase of the stellar mass *M*) with $\sigma$ up to some maximum values (say, $\sigma_{CR}$). It is worth mention that the critical value $\sigma_{CR}$ marks the onset of the first mode of radial instability. For the case $n = 1.0$, Fig. 1 shows that the critical value $\sigma_{CR} = 0.42$ and the relativistic polytrope are stable for $\sigma < 0.42$. In figures 2, 3, and 4 we observe critical values of the general relativity index $\sigma_{CR} = 0.2$, $\sigma_{CR} = 0.1$ and $\sigma_{CR} = 0.04$ for the cases $n = 1.5$, $n = 2.0$ and $n = 2.5$ respectively. In Fig. 5 where $n = 3.0$, $\tilde{M}$ has a maximum at $\sigma_{CR} = 0$ which marks the onset of the first mode of instability, while the minimum value at $\sigma_{CR} = 0.53$ marks the onset of the next mode of nonradial instability. In this case, Equation 17 reduces to $\tilde{M} \equiv \upsilon(\xi_1)$. It is concluded that for $\sigma_{CR} > 0.5$ the relativistic polytropic models are energetically unstable.

The study of the stability of polytropes is useful for determining some physical properties such as the maximum mass limit and illustrates how the stellar mass increases or decreases due to the effects of general relativity. For a given mass, radius, and a polytropic index *n*, Fig. 6 of mass-radius relation can determine the internal structure of a polytrope. This means each value of a relativistic parameter $\sigma$ corresponds to a certain internal structure. We can see from Fig. 6 that one pair of mass and radius has two different values of $\sigma$. For the case of a polytropic index n=3.0, the logarithmic function $\log_{10}[\sigma\ (n+1)\upsilon(\xi_1)/\xi_1)] = -2.03$ has two values of $\sigma \simeq 0.67$ and $\sigma \simeq 0.75$. Then we have two spherical polytropic configurations of the same mass and radius but different internal structures. When n=2.0, the logarithmic function $\log_{10}[\sigma\ (n+1)\upsilon(\xi_1)/\xi_1)] = -0.76$ has two values of $\sigma \simeq 0.42$ and $\sigma \simeq 0.47$. Such information reflects the importance of relativistic solutions.

Table 1, gives the limits of mass-radius; for example, if the polytropic index n=1.0 $GM/c^2\overline{R} \leq 0.214$, then the gravitational radius $2GM/c^2$ is at most 43% of the invariant(physical) radius $\overline{R}$. When the polytropic index n=3.0 $GM/c^2\overline{R} \leq 0.072$, then the gravitational radius $2GM/c^2$ is at most 14.5% of the invariant radius $\overline{R}$, which is very small compared to the limit value when n=1.0.

The results of all critical values obtained in this paper for different polytropic indices are in full agreement with those evaluated by several authors such as Tooper(1964), Bludman(1973), and Araujo & Chirenti (2011). These critical values $\sigma_{CR}$ and $\tilde{M}(\sigma)$ together with various indices *n* are given in Table A6 (Appendix A). It is shown that the spherical polytrope of index n=3.0 and $\sigma > 0.5$ is energetically unstable.

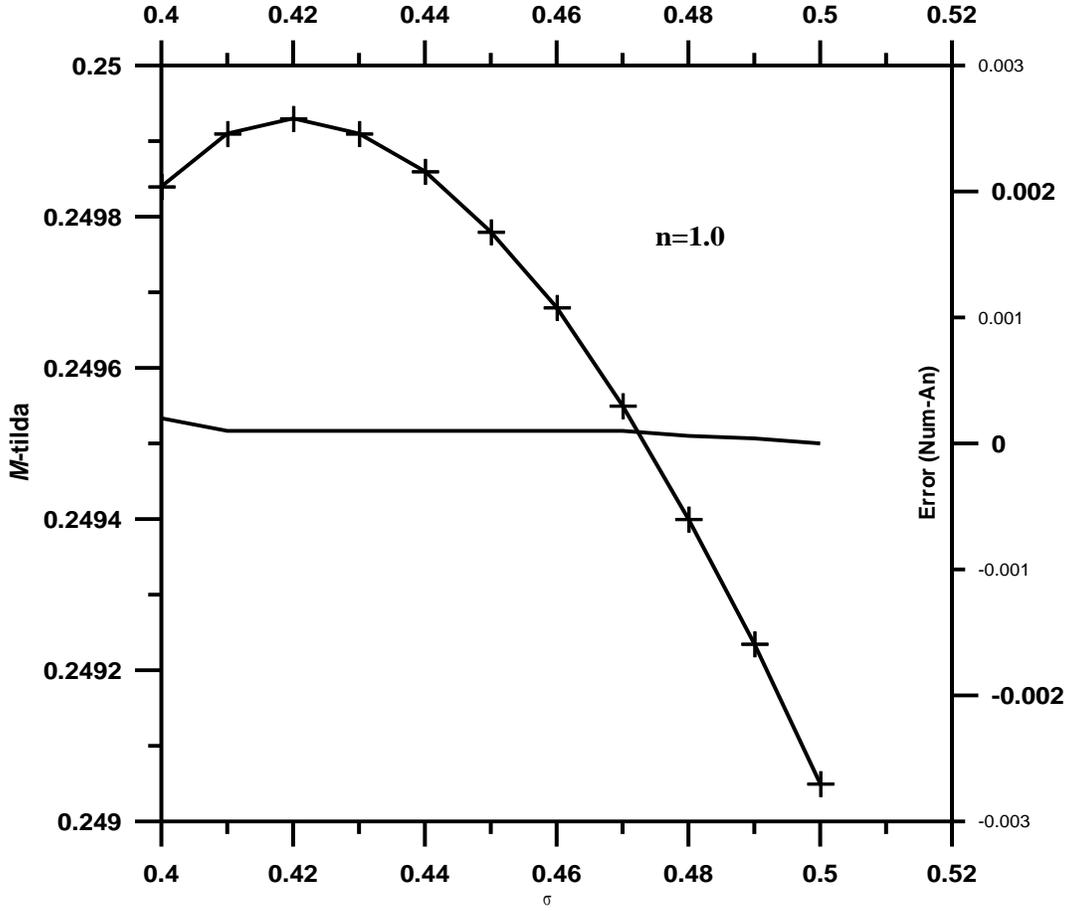

Fig. 1. $\tilde{M}(\sigma)$ for $n=1$, and $\sigma_{CR} \approx 0.42$. Stable relativistic polytrope occurs for $\sigma < 0.42$, while unstable models occur when $\sigma > 0.42$. Comparison of analytical and numerical results provides errors of order $10^{-4}$.

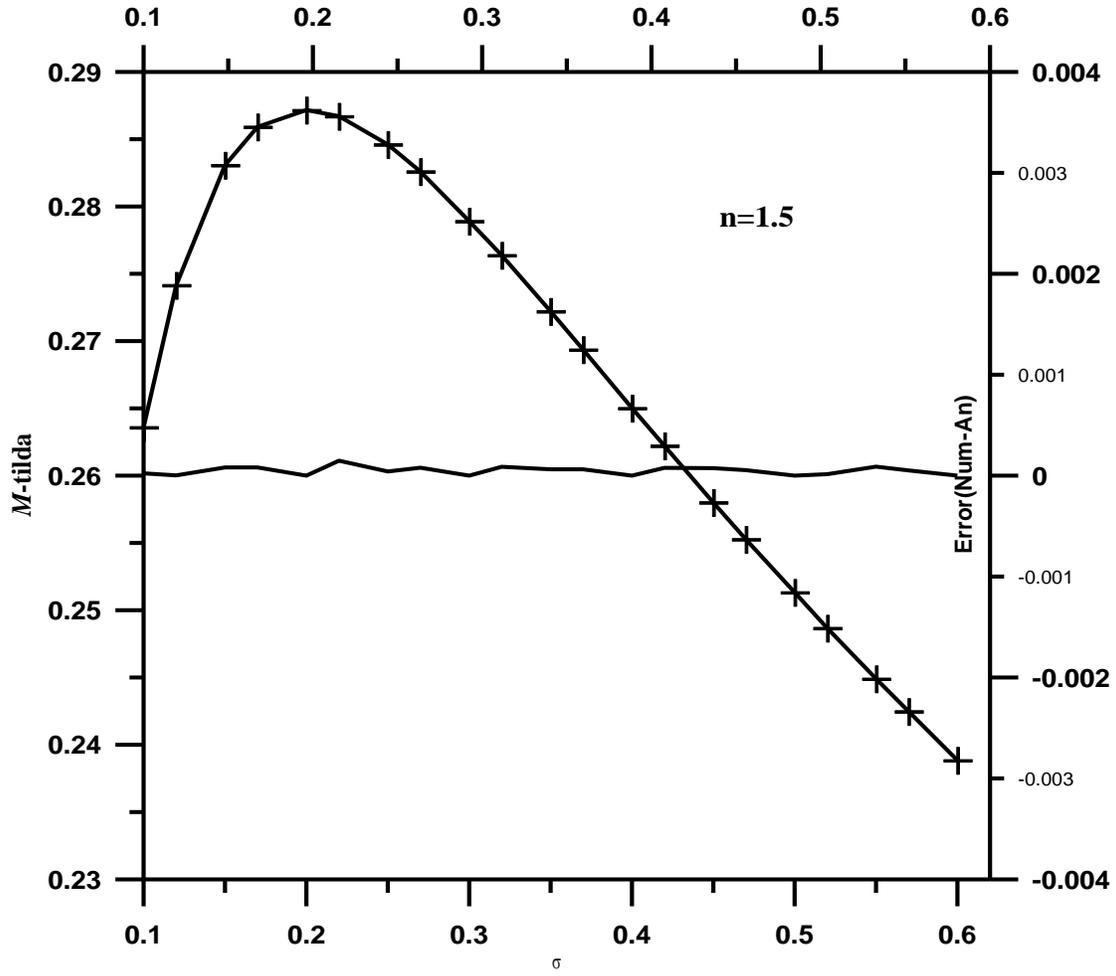

Fig. 2. $\tilde{M}(\sigma)$ for $n = 1.5$, and $\sigma_{CR} \approx 0.2$. Stable relativistic polytrope occurs for $\sigma < 0.2$, while unstable models occur when $\sigma > 0.2$. Comparison of analytical and numerical results provides errors of order $10^{-5}$.

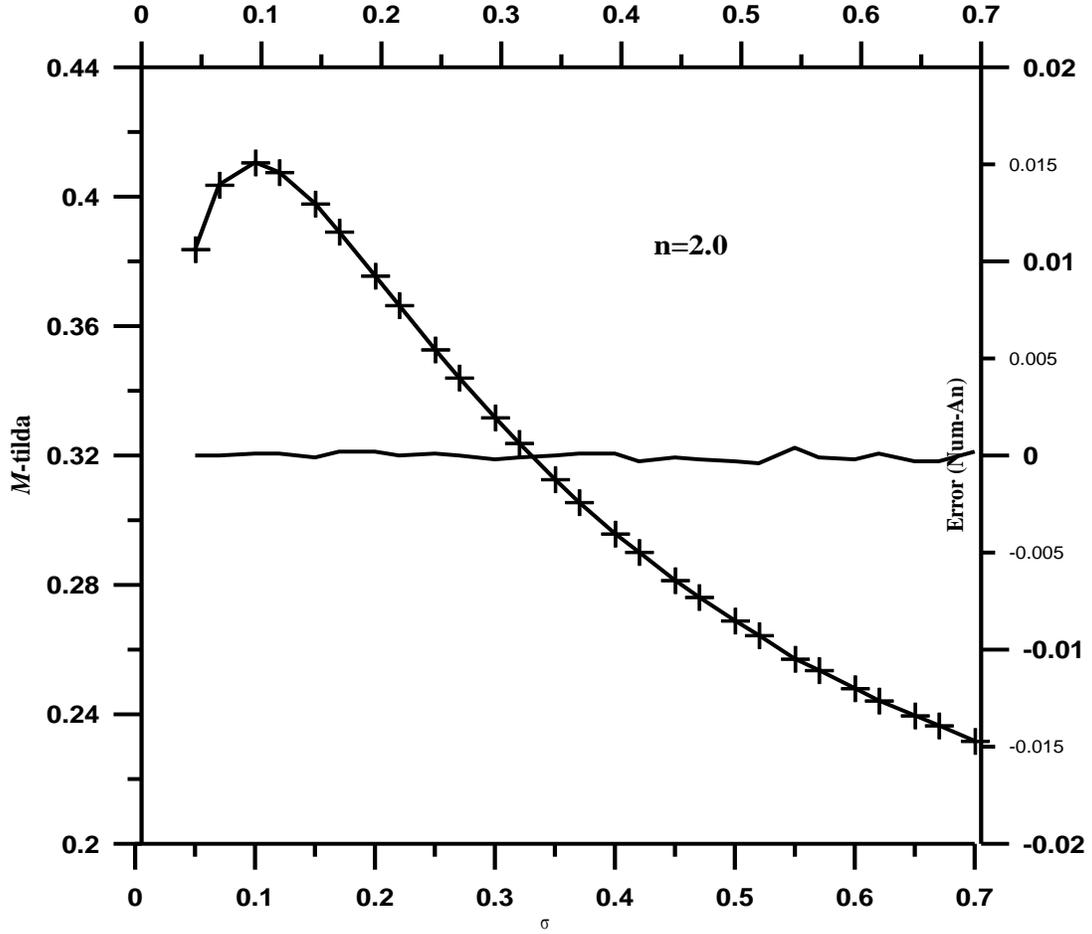

Fig. 3. $\tilde{M}(\sigma)$ for $n=2$, and $\sigma_{CR} \approx 0.1$. Stable relativistic polytrope occurs for $\sigma < 0.1$, while unstable models occur when $\sigma > 0.1$. Comparison of analytical and numerical results provides errors of order $10^{-4}$.

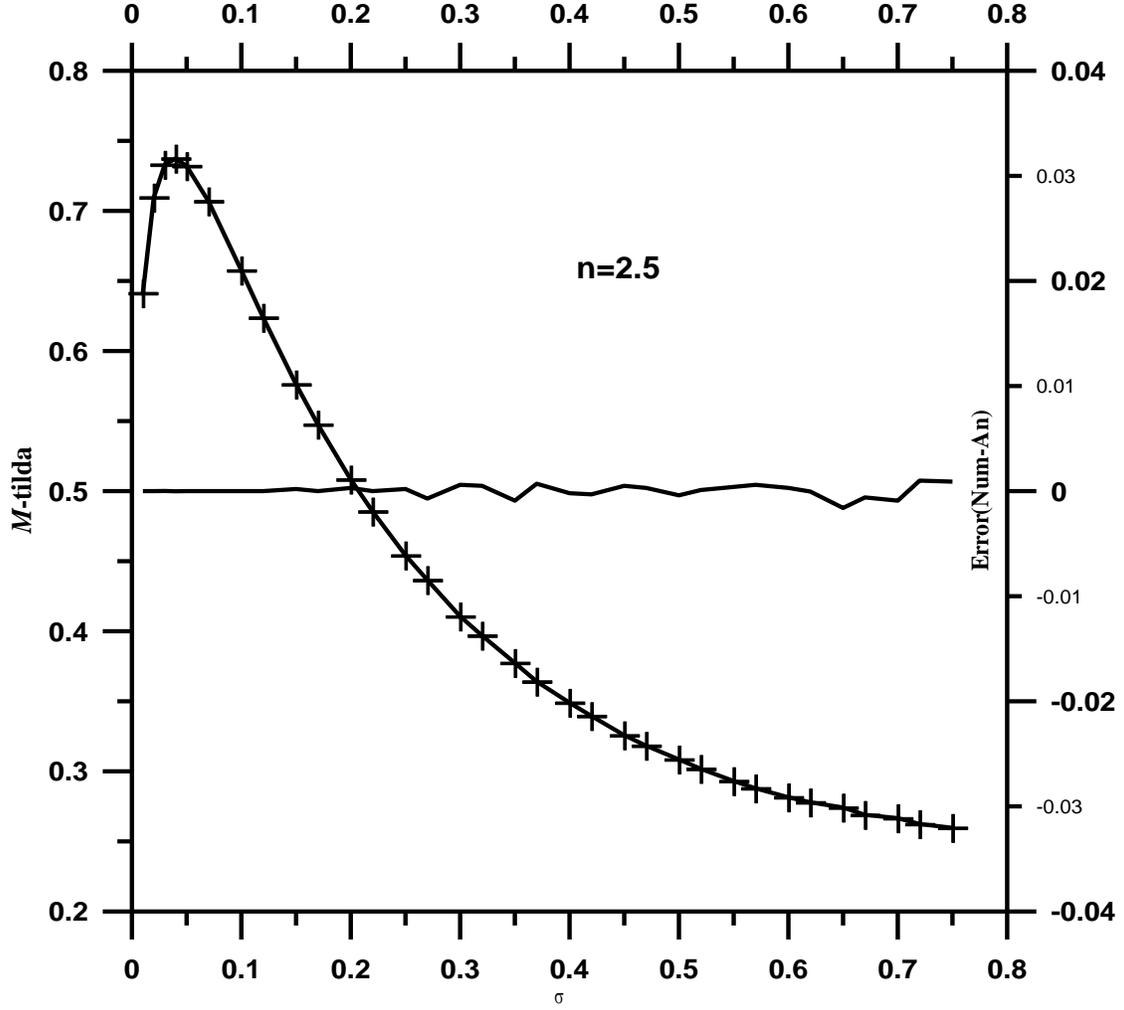

Fig. 4. $\tilde{M}(\sigma)$ for $n = 2.5$, and $\sigma_{CR} \approx 0.04$. Stable relativistic polytrope occurs for $\sigma < 0.04$, while unstable models occur when $\sigma > 0.04$. Comparison of analytical and numerical results provides errors of order $10^{-4}$.

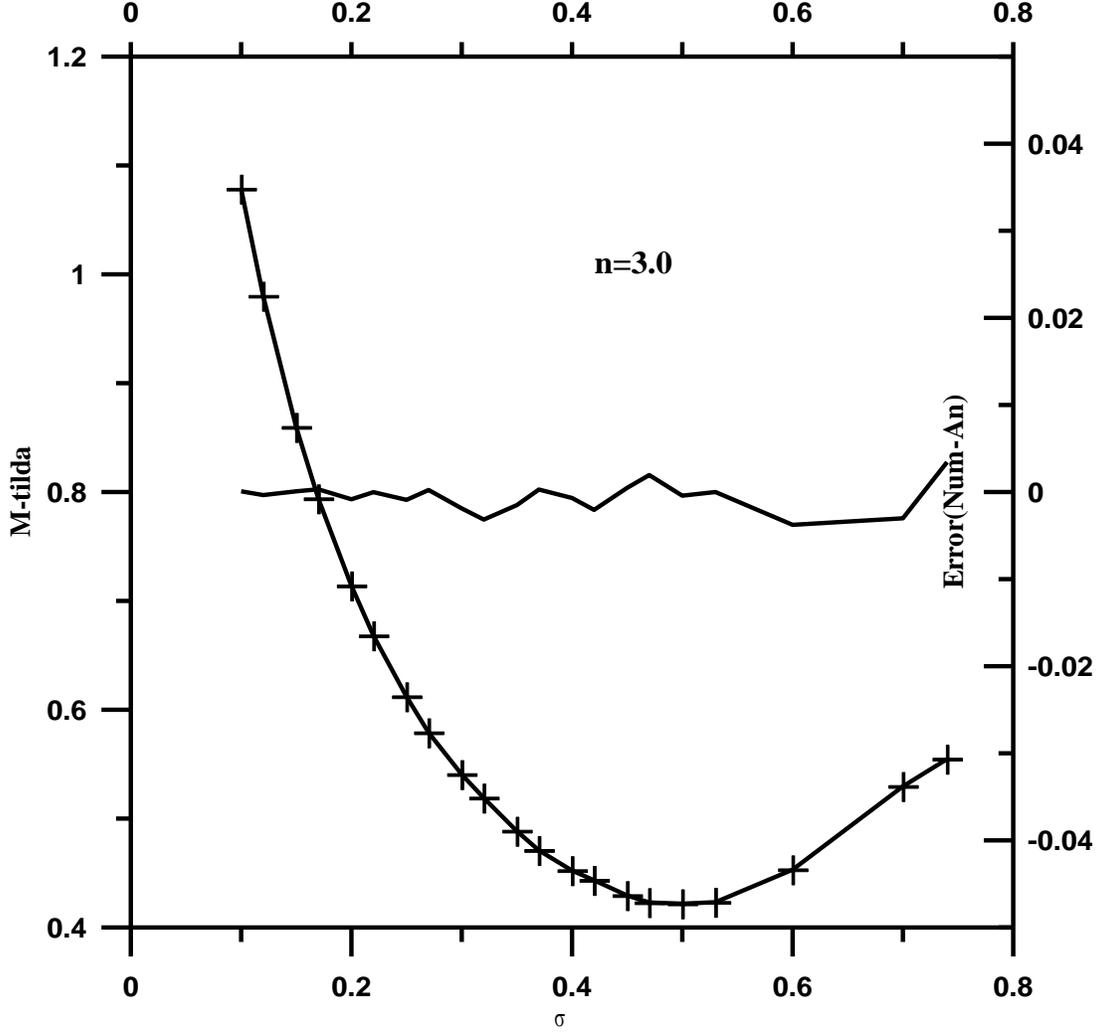

Fig. 5 for $n = 3.0$, $\tilde{M}$ has a maximum at $\sigma_{CR} = 0$ which marks the onset of the first mode of instability, while the minimum at $\sigma_{CR} \approx 0.53$ marks the onset of the next mode of nonradial instability. Comparison of analytical and numerical results provides errors of order $10^{-3}$.

The mass-radius relation (Tooper 1964) has the form:

$$\frac{GM}{c^2 \overline{R}} = \frac{\sigma \ (n+1) \upsilon(\xi_1)}{\overline{\xi}_1}, \tag{20}$$

where $\overline{R}$ defines the physical radius (invariant radius) of the sphere and $\overline{\xi}_1 = A\overline{R}$ can be obtained by integrating the Equation

$$\overline{\xi}_1 = \int_0^{\xi_1} \left(1 - 2\sigma \ (n+1)\upsilon(\xi)/\xi\right)^{-1/2} d\xi. \tag{21}$$

The mass-radius relation is useful for determining the surface redshift. It gives the ratio of the gravitational radius $2GM/c^2$ to the invariant radius $\overline{R}$ when $n$ and $\sigma$ are known. Rewrite Equation (20) in terms of numerical values for solar mass and solar radius and take logarithms of both sides of the resulting equation (Tooper 1964). Then using the solutions introduced in section 3, we plotted the logarithmic ratio of gravitational radius to a geometrical radius as a function of the relativistic parameter $\sigma$ for different values of the polytropic index $n$ [see Fig. 6]. Table 1 gives the limits of mass-radius relations for different polytropic index n.

Table 1: Limits of mass-radius relations

| n | Max. value of $\log_{10}[\sigma\,(n+1)\upsilon(\xi_1)/\overline{\xi_1}]$ | Limit ratio of $GM/c^2\overline{R}$ | Limit ratio of $GM/c^2 R$ |
|---|---|---|---|
| 1.0 | -0.670 | 0.214 | 0.277 |
| 1.5 | -0.769 | 0.170 | 0.237 |
| 2.0 | -0.885 | 0.130 | 0.174 |
| 2.5 | -1.022 | 0.095 | 0.117 |
| 3.0 | -1.201 | 0.0633 | 0.072 |

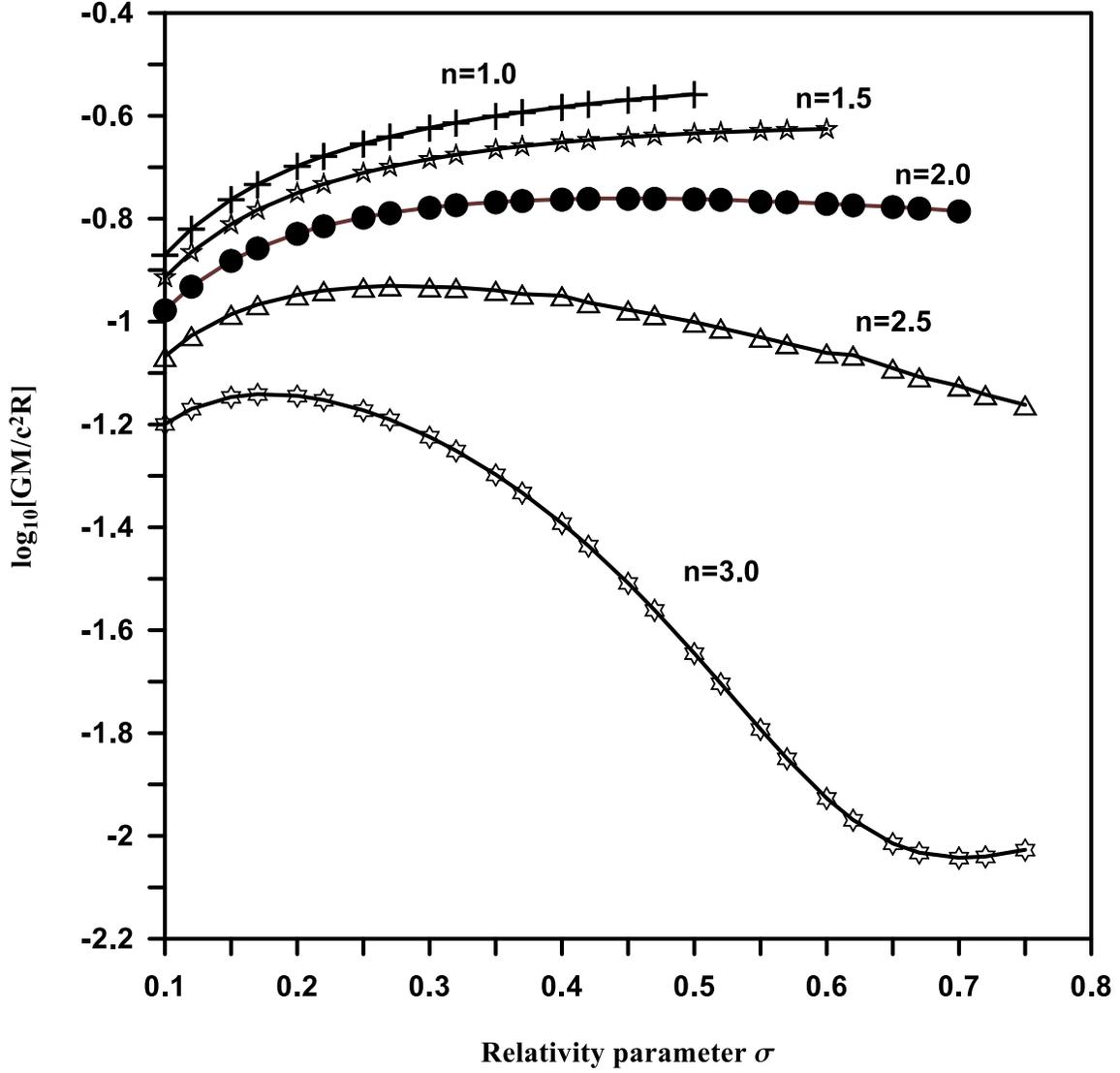

Fig. 6. The logarithmic ratio of the gravitational radius to the coordinate radius, as a function of the relativistic parameter, $\sigma$, for different values of the polytropic index $n$.

## 5. Series Convergence

The power series solution of the relativistic problem without using any acceleration techniques is very limited. Tables 2, 3, and 4 show the radius of convergence $\xi_1$ of the power series solution $\theta(\xi_1)$ of Equation (13) and the relative error ($\varepsilon$) before performing any acceleration. When the polytropic indices n=1.0 and n=1.5, the series is convergent rapidly. However, beyond the mentioned values of the polytropic index, the power series solution is either slowly convergent or divergent. It is noted that the relative error ($\varepsilon = |\xi_1(An) - \xi_1(Num)| / \xi_1(Num)$) increases gradually with increasing the polytropic effects $\sigma$ and the polytropic index n as well. Which in turn results in a small physical range of the convergent power series solutions. Hence produces inaccurate physical parameters of the relativistic polytropes.

The fourth order Runge-Kutta method was used for the performance of the numerical solution of the relativistic TOV equation. Analytical and numerical calculations through this paper have been done using the Mathematica package, version 11.2.

To extend the physical radii of the convergent power series solutions, a combination of the two techniques for Euler-Abel transformation and Padé approximation Nouh (2004) and Nouh & Saad(2013) have been utilized. Tables from A1 to A5 (Appendix A) show comparisons between numerical and analytical results. It is worth noting that the power series solutions are rapidly convergent for the polytropic indices n=1(0.5)-3.0 and provide a maximum relative error of order $10^{-3}$.

**Table 2. Radii of the convergence of $\theta(\xi)$ and relative error for n=1.0**

| $\sigma$ | $\xi_1(N)$ | $\xi_1(A)$ | $\varepsilon$ : relative error |
|---|---|---|---|
| 0.1 | 2.5990 | 2.5990 | 0.0 |
| 0.2 | 2.2770 | 2.2765 | 0.000219635 |
| 0.3 | 2.0641 | 2.0637 | 0.000193827 |
| 0.4 | 1.9132 | 1.9111 | 0.001098844 |
| 0.5 | 1.8008 | 1.8862 | 0.045276217 |

**Table 3. Radii of the convergence of $\theta(\xi)$ and relative error for n=1.5**

| $\sigma$ | $\xi_1(N)$ | $\xi_1(A)$ | $\varepsilon$ : relative error |
|---|---|---|---|
| 0.1 | 3.0384 | 3.0730 | 0.011259356 |
| 0.2 | 2.6993 | 2.6025 | 0.037195005 |
| 0.3 | 2.4930 | 2.4281 | 0.026728718 |
| 0.4 | 2.3610 | 2.2648 | 0.042476157 |
| 0.5 | 2.2749 | 2.0644 | 0.101966673 |
| 0.6 | 2.2192 | 1.8340 | 0.210032715 |

**Table 4. Radii of the convergence of $\theta(\xi)$ and relative error for n=2.0**

| $\sigma$ | $\xi_1(N)$ | $\xi_1(A)$ | $\varepsilon$: relative error |
|---|---|---|---|
| 0.1 | 3.6989 | 3.4259 | 0.07968709 |
| 0.2 | 3.3983 | 2.5632 | 0.325803683 |
| 0.3 | 3.2711 | 2.5577 | 0.278922469 |
| 0.4 | 3.2473 | 1.9503 | 0.665025893 |
| 0.5 | 3.2967 | 1.9836 | 0.661978221 |
| 0.6 | 3.3986 | 1.7686 | 0.92163293 |
| 0.67 | 3.4982 | 1.6556 | 1.112949988 |

## 6. Conclusion

In the present paper, we considered the stability properties of the relativistic polytrope. We analyze for various polytropic indices the stability of the relativistic polytrope. An analytic solution is applied to the TOV equation that provides us with relativistic polytropic physical parameters. For each polytropic index, we tested the critical values of the relativistic parameter at which the radial instability started. It is shown that for a given mass, radius and a polytropic index *n*, the internal structure of a polytropic fluid sphere can be determined as a function of the relativistic parameter $\sigma$. For n=1 (0.5)-2.5, stable relativistic polytropes occur for $\sigma$ less than the critical values 0.42, 0.20, 0.10, and 0.04 respectively, while unstable relativistic polytropes are obtained when the relativistic parameter $\sigma$ is greater than the same values. When n=3.0 and $\sigma > 0.5$, energetically unstable solutions have occurred.

# Appendix A: Numerical Results

In the following tables, we listed the numerical results obtained for different polytropic indices. The designation of the columns are as follows:

1- $\sigma$ is the relativistic parameter.

2- $\xi 1$ is the first zero of the Emden function.

3- $v(\xi 1)$Num: numerical solution of the relativistic function.

4- $v(\xi 1)$An: analytical solution of the relativistic function.

5- $\tilde{M}(\sigma)$ Num is a parameter analog to the mass of the polytrope computed numerically.

6- $\tilde{M}(\sigma)$ An is a parameter analog to the mass of the polytrope computed analytically.

7- $\Delta v(\xi 1)$ is the difference between the analytical and the numerical values of the function $v$.

8- $\Delta \tilde{M}(\sigma)$ is the difference between the analytical and the numerical values.

9- $\sigma_{critical}$ is the critical value of the fractional parameter at which instability started.

Table A1. Comparisons between analytical and numerical solutions of the relativistic functions $v(\xi 1)$ and $\tilde{M}(\sigma)$ for n=1.0

| $\sigma$ | $\xi 1$ | $v(\xi_1)$Num | $v(\xi 1)$An | $\tilde{M}(\sigma)$ Num | $\tilde{M}(\sigma)$ An | $\Delta v(\xi 1)$ | $\Delta \tilde{M}(\sigma)$ |
|---|---|---|---|---|---|---|---|
| 0.0 | 3.1415 | 3.1416 | 3.1416 | 0.0 | 0.0 | 0.0 | 0.0 |
| 0.10 | 2.5990 | 1.7514 | 1.7514 | 0.1751 | 0.1751 | 0.0 | 0.0 |
| 0.12 | 2.5221 | 1.5922 | 1.5922 | 0.1911 | 0.1911 | 0.0 | 0.0 |
| 0.15 | 2.4198 | 1.3941 | 1.3941 | 0.2091 | 0.2091 | 0.0 | 0.0 |
| 0.17 | 2.3590 | 1.2834 | 1.2835 | 0.2182 | 0.2182 | 0.0001 | 0.0 |
| 0.20 | 2.2770 | 1.1426 | 1.1426 | 0.2285 | 0.2285 | 0.0 | 0.0 |
| 0.22 | 2.2278 | 1.0624 | 1.0624 | 0.2337 | 0.2337 | 0.0 | 0.0 |
| 0.25 | 2.1607 | 0.9585 | 0.9585 | 0.2396 | 0.2396 | 0.0 | 0.0 |
| 0.27 | 2.1200 | 0.8983 | 0.8981 | 0.2425 | 0.2425 | 0.0002 | 0.0 |
| 0.30 | 2.0641 | 0.8192 | 0.8190 | 0.2457 | 0.2457 | 0.0002 | 0.0 |
| 0.32 | 2.0299 | 0.7727 | 0.7728 | 0.2473 | 0.2473 | -0.0001 | 0.0 |
| 0.35 | 1.9827 | 0.7109 | 0.7109 | 0.2488 | 0.2488 | 0.0 | 0.0 |
| 0.37 | 1.9536 | 0.6742 | 0.6742 | 0.2495 | 0.2495 | 0.0 | 0.0 |
| 0.40 | 1.9132 | 0.6249 | 0.6250 | 0.2500 | 0.2500 | 0.0001 | 0.0 |
| 0.42 | 1.8882 | 0.5954 | 0.5954 | 0.2501 | 0.2501 | 0.0 | 0.0 |
| 0.45 | 1.8531 | 0.5553 | 0.5554 | 0.2499 | 0.2499 | 0.0001 | 0.0 |
| 0.47 | 1.8314 | 0.5311 | 0.5310 | 0.2496 | 0.2496 | 0.0001 | 0.0 |
| 0.50 | 1.8008 | 0.4981 | 0.4981 | 0.2491 | 0.2491 | 0.0 | 0.0 |

Table A2. Comparisons between analytical and numerical solutions of the relativistic functions
$v(\xi_1)$ and $\tilde{M}(\sigma)$ for n=1.5

| $\sigma$ | $\xi_1$ | $v(\xi_1)$Num | $v(\xi_1)$ An | $\tilde{M}(\sigma)$ Num | $\tilde{M}(\sigma)$ An | $\Delta\tilde{M}(\sigma)$ |
|---|---|---|---|---|---|---|
| 0.0 | 3.6537 | 2.7141 | 2.7141 | 0.0 | 0.0 | 0.0 |
| 0.10 | 3.0384 | 1.4823 | 1.4822 | 0.263592 | 0.263569 | 2.3E-05 |
| 0.12 | 2.9552 | 1.3446 | 1.3446 | 0.274153 | 0.274151 | 2E-06 |
| 0.15 | 2.8464 | 1.1744 | 1.1741 | 0.283069 | 0.282987 | 8.2E-05 |
| 0.17 | 2.783 | 1.08 | 1.08 | 0.285922 | 0.285841 | 8.1E-05 |
| 0.20 | 2.6993 | 0.9602 | 0.9602 | 0.287166 | 0.287166 | 0.0 |
| 0.22 | 2.65 | 0.8925 | 0.8921 | 0.286706 | 0.286558 | 0.000148 |
| 0.25 | 2.5843 | 0.805 | 0.8049 | 0.284609 | 0.284568 | 4.1E-05 |
| 0.27 | 2.5453 | 0.7545 | 0.7543 | 0.282603 | 0.282524 | 7.9E-05 |
| 0.30 | 2.493 | 0.6881 | 0.6881 | 0.278913 | 0.278913 | 0.0 |
| 0.32 | 2.4619 | 0.6496 | 0.6494 | 0.27638 | 0.276292 | 8.8E-05 |
| 0.35 | 2.42 | 0.5982 | 0.5981 | 0.272214 | 0.272151 | 6.3E-05 |
| 0.37 | 2.3949 | 0.5678 | 0.5677 | 0.269366 | 0.269303 | 6.3E-05 |
| 0.40 | 2.361 | 0.5269 | 0.5269 | 0.265018 | 0.265018 | 0.0 |
| 0.42 | 2.3407 | 0.5026 | 0.5025 | 0.262221 | 0.262144 | 7.7E-05 |
| 0.45 | 2.3134 | 0.4696 | 0.4695 | 0.258014 | 0.25794 | 7.4E-05 |
| 0.47 | 2.297 | 0.4497 | 0.4496 | 0.25527 | 0.255216 | 5.4E-05 |
| 0.50 | 2.2749 | 0.4227 | 0.4227 | 0.251326 | 0.251326 | 0.0 |
| 0.52 | 2.2617 | 0.4061 | 0.4061 | 0.248669 | 0.248655 | 1.4E-05 |
| 0.55 | 2.2439 | 0.3835 | 0.3833 | 0.244906 | 0.244817 | 8.9E-05 |
| 0.57 | 2.2333 | 0.3696 | 0.3696 | 0.242484 | 0.242432 | 5.2E-05 |
| 0.60 | 2.2192 | 0.3504 | 0.3504 | 0.238846 | 0.238846 | 0.0 |

Table A3. Comparisons between analytical and numerical solutions of the relativistic functions
$v(\xi_1)$ and $\tilde{M}(\sigma)$ for n=2.0

| $\sigma$ | $\xi_1$ | $v(\xi_1)$Num | $v(\xi_1)$An | $\tilde{M}(\sigma)$ Num | $\tilde{M}(\sigma)$ An | $\Delta\tilde{M}(\sigma)$ |
|---|---|---|---|---|---|---|
| 0.0 | 4.3531 | 2.411 | 2.411 | 0.0 | 0.0 | 0.0 |
| 0.05 | 3.9617 | 1.7165 | 1.7162 | 0.3838 | 0.3838 | 0.0 |
| 0.07 | 3.8443 | 1.5258 | 1.5258 | 0.4037 | 0.4037 | 0.0 |
| 0.10 | 3.6989 | 1.2987 | 1.2983 | 0.4107 | 0.4106 | 0.0001 |
| 0.12 | 3.6191 | 1.1769 | 1.1766 | 0.4077 | 0.4076 | 0.0001 |
| 0.15 | 3.5198 | 1.0272 | 1.0274 | 0.3978 | 0.3979 | -0.0001 |
| 0.17 | 3.4653 | 0.9445 | 0.9440 | 0.3894 | 0.3892 | 0.0002 |
| 0.20 | 3.3983 | 0.8403 | 0.8399 | 0.3758 | 0.3756 | 0.0002 |
| 0.22 | 3.3619 | 0.7814 | 0.7815 | 0.3665 | 0.3665 | 0.0 |
| 0.25 | 3.3186 | 0.7058 | 0.7056 | 0.3529 | 0.3528 | 0.0001 |
| 0.27 | 3.2962 | 0.6623 | 0.6622 | 0.3441 | 0.3441 | 0.0 |
| 0.30 | 3.2711 | 0.6055 | 0.6057 | 0.3316 | 0.3318 | -0.0002 |
| 0.32 | 3.2595 | 0.5723 | 0.5725 | 0.3238 | 0.3239 | -0.0001 |
| 0.35 | 3.2491 | 0.5285 | 0.5285 | 0.3127 | 0.3127 | 0.0 |
| 0.37 | 3.2463 | 0.5026 | 0.5024 | 0.3057 | 0.3056 | 0.0001 |
| 0.40 | 3.2473 | 0.4680 | 0.4678 | 0.2960 | 0.2959 | 0.0001 |
| 0.42 | 3.2526 | 0.4474 | 0.4478 | 0.2899 | 0.2902 | -0.0003 |
| 0.45 | 3.2644 | 0.4195 | 0.4196 | 0.2814 | 0.2815 | -0.0001 |
| 0.47 | 3.2754 | 0.4028 | 0.4030 | 0.2761 | 0.2763 | -0.0002 |
| 0.50 | 3.2967 | 0.3800 | 0.3804 | 0.2687 | 0.2690 | -0.0003 |
| 0.52 | 3.3128 | 0.3662 | 0.3668 | 0.2641 | 0.2645 | -0.0004 |
| 0.55 | 3.3416 | 0.3474 | 0.3468 | 0.2576 | 0.2572 | 0.0004 |
| 0.57 | 3.3632 | 0.3359 | 0.3360 | 0.2536 | 0.2537 | -0.0001 |
| 0.60 | 3.3986 | 0.3201 | 0.3202 | 0.2479 | 0.2481 | -0.0002 |
| 0.62 | 3.4253 | 0.3104 | 0.3103 | 0.2444 | 0.2443 | 0.0001 |
| 0.65 | 3.4678 | 0.2970 | 0.2973 | 0.2394 | 0.2397 | -0.0003 |
| 0.67 | 3.4982 | 0.2887 | 0.2891 | 0.2363 | 0.2366 | -0.0003 |

Table A4. Comparisons between analytical and numerical solutions of the relativistic functions $v(\xi_1)$ and $\tilde{M}(\sigma)$ for n=2.5

| $\sigma$ | $\xi_1$ | $v(\xi_1)$Num | $v(\xi_1)$An | $\tilde{M}(\sigma)$ Num | $\tilde{M}(\sigma)$ An | $\Delta\tilde{M}(\sigma)$ |
|---|---|---|---|---|---|---|
| 0.0 | 5.3552 | 2.1872 | 2.1872 | 0.0 | 0.0 | 0.0 |
| 0.01 | 5.2623 | 2.0281 | 2.0281 | 0.641341 | 0.641341 | 0.0 |
| 0.02 | 5.1793 | 1.88702 | 1.88702 | 0.709633 | 0.709632 | 0.0 |
| 0.03 | 5.1052 | 1.76134 | 1.76131 | 0.733034 | 0.7033021 | 1.3E-05 |
| 0.04 | 5.0393 | 1.648899 | 1.648930 | 0.737410 | 0.737424 | -1.4E-05 |
| 0.05 | 4.9809 | 1.5479 | 1.548019 | 0.7320 | 0.732013 | 0.0 |
| 0.07 | 4.8841 | 1.3744 | 1.374359 | 0.7069 | 0.706927 | 0.0 |
| 0.10 | 4.7819 | 1.1692 | 1.169269 | 0.6575 | 0.657528 | 0.0 |
| 0.12 | 4.7383 | 1.0599 | 1.059908 | 0.6238 | 0.623826 | 0.0 |
| 0.15 | 4.7044 | 0.9261 | 0.925878 | 0.5764 | 0.576204 | 0.0002 |
| 0.17 | 4.7006 | 0.8527 | 0.852617 | 0.5475 | 0.547477 | 0.0 |
| 0.20 | 4.7206 | 0.7606 | 0.760086 | 0.5086 | 0.508299 | 0.0003 |
| 0.22 | 4.7498 | 0.7088 | 0.708902 | 0.4855 | 0.485503 | 0.0 |
| 0.25 | 4.8163 | 0.6426 | 0.642273 | 0.4544 | 0.454155 | 0.0002 |
| 0.27 | 4.8753 | 0.6048 | 0.605615 | 0.4359 | 0.436554 | -0.0007 |
| 0.30 | 4.9855 | 0.5556 | 0.554771 | 0.4112 | 0.410576 | 0.0006 |
| 0.32 | 5.0734 | 0.5271 | 0.527703 | 0.3964 | 0.396896 | 0.0005 |
| 0.35 | 5.2273 | 0.4896 | 0.490730 | 0.3766 | 0.377450 | -0.0009 |
| 0.37 | 5.3450 | 0.4677 | 0.466891 | 0.3648 | 0.364138 | 0.0007 |
| 0.40 | 5.5448 | 0.438571 | 0.444662 | 0.348782 | 0.348982 | -0.0002 |
| 0.42 | 5.6943 | 0.4214 | 0.421721 | 0.3392 | 0.339499 | -0.0003 |
| 0.45 | 5.9440 | 0.3984 | 0.397734 | 0.3263 | 0.325758 | 0.0005 |
| 0.47 | 6.1284 | 0.3847 | 0.384457 | 0.3186 | 0.318326 | 0.0003 |
| 0.50 | 6.4335 | 0.3664 | 0.366915 | 0.3081 | 0.308537 | -0.0004 |
| 0.52 | 6.6569 | 0.3555 | 0.355406 | 0.3019 | 0.301804 | 0.0001 |
| 0.55 | 7.0239 | 0.3408 | 0.340363 | 0.2935 | 0.293112 | 0.0004 |
| 0.57 | 7.2910 | 0.3320 | 0.331352 | 0.2885 | 0.287911 | 0.0006 |
| 0.60 | 7.7273 | 0.3202 | 0.319843 | 0.2818 | 0.281497 | 0.0003 |
| 0.62 | 8.0423 | 0.3131 | 0.319070 | 0.277851 | 0.277888 | -3.7E-05 |
| 0.65 | 8.5563 | 0.3036 | 0.305345 | 0.2726 | 0.274169 | -0.0016 |
| 0.67 | 8.9257 | 0.2980 | 0.297274 | 0.2696 | 0.268953 | -0.0006 |
| 0.70 | 9.5224 | 0.2905 | 0.291442 | 0.2657 | 0.266579 | -0.0009 |
| 0.72 | 9.9494 | 0.2860 | 0.284817 | 0.2634 | 0.2623610 | 0.001 |

Table A5. Comparisons between analytical and numerical solutions of the relativistic functions $v(\xi_1)$ and $\tilde{M}(\sigma)$ for n=3.0

| $\sigma$ | $\xi_1$ | $v(\xi_1)$Num | $v(\xi_1)$An | $\tilde{M}(\sigma)$ Num | $\tilde{M}(\sigma)$ An | $\Delta\tilde{M}(\sigma)$ |
|---|---|---|---|---|---|---|
| 0.0 | 6.8968 | 2.01824 | 2.01824 | 2.01824 | 2.01824 | 0.0 |
| 0.05 | 6.7074 | 1.42463 | 1.42463 | 1.42463 | 1.42463 | 0.0 |
| 0.07 | 6.7206 | 1.26543 | 1.26542 | 1.26543 | 1.26542 | 1.0E-05 |
| 0.10 | 6.8258 | 1.07845 | 1.07837 | 1.07845 | 1.07837 | 8.0E-05 |
| 0.12 | 6.9521 | 0.979601 | 0.979949 | 0.979601 | 0.979949 | - 0.0003 |
| 0.15 | 7.2285 | 0.85958 | 0.859491 | 0.85958 | 0.859491 | 9.0E-05 |
| 0.17 | 7.4751 | 0.794229 | 0.793908 | 0.794229 | 0.793908 | 0.0003 |
| 0.20 | 7.9508 | 0.713042 | 0.713880 | 0.713042 | 0.713880 | -0.0008 |
| 0.22 | 8.3481 | 0.667954 | 0.667963 | 0.667954 | 0.667963 | -1.0E-05 |
| 0.25 | 9.0894 | 0.611096 | 0.612004 | 0.611096 | 0.612004 | -0.0009 |
| 0.27 | 9.6994 | 0.579159 | 0.578934 | 0.579159 | 0.578934 | 0.0002 |
| 0.30 | 10.8327 | 0.538631 | 0.540522 | 0.538631 | 0.540522 | -0.0018 |
| 0.32 | 11.7690 | 0.515833 | 0.519000 | 0.515833 | 0.519000 | -0.003 |
| 0.35 | 13.5271 | 0.487068 | 0.488563 | 0.487068 | 0.488563 | -0.001 |
| 0.37 | 15.0007 | 0.471124 | 0.470841 | 0.471124 | 0.470841 | 0.0003 |
| 0.40 | 17.8197 | 0.451585 | 0.452268 | 0.451585 | 0.452268 | -0.0007 |
| 0.42 | 20.2306 | 0.441299 | 0.443341 | 0.441299 | 0.443341 | -0.002 |
| 0.45 | 24.9438 | 0.429831 | 0.429350 | 0.429831 | 0.429350 | 0.0005 |
| 0.47 | 29.0538 | 0.424822 | 0.422867 | 0.424822 | 0.422867 | 0.0020 |
| 0.50 | 37.2058 | 0.421395 | 0.421807 | 0.421395 | 0.421807 | -0.0004 |
| 0.53 | 48.5317 | 0.423168 | 0.418075 | 0.423168 | 0.418075 | 0.0051 |
| 0.60 | 91.0723 | 0.449319 | 0.453089 | 0.449319 | 0.453089 | -0.004 |
| 0.70 | 162.5832 | 0.526621 | 0.529641 | 0.526621 | 0.529641 | -0.003 |
| 0.74 | 177.9357 | 0.558153 | 0.554724 | 0.558153 | 0.554724 | 0.003 |
| 0.75 | 180.4379 | 0.565394 | 0.541169 | 0.565394 | 0.541169 | 0.0242 |

Table A6. The critical values $\sigma_{CR}$ corresponding $\tilde{M}(\sigma)$ for various indices $n$

| n | $\xi_1$ | $\sigma_{critical}$ | $\tilde{M}(\sigma)$ |
|---|---|---|---|
| 1.0 | 1.8882 | 0.42 | 0.249930 |
| 1.5 | 2.6993 | 0.20 | 0.287166 |
| 2.0 | 3.6989 | 0.10 | 0.410546 |
| 2.5 | 5.0393 | 0.04 | 0.737424 |
| 3.0 | 6.8968 | 0.00 | 2.01824 |
| 3.0 | 48.5317 | 0.53 | 0.416203 |

## Appendix B: The Series Acceleration Technique

To accelerate the convergence of the series solution of Equation (13), we followed the scheme developed by Nouh (2004). As the first step of this scheme, the alternating series is accelerated by Euler–Abel transformation (Demodovich and Maron, 1973).

Let us write

$$\theta(\xi) = a_0 + \xi\, \phi(\xi), \tag{1}$$

Where

$$\phi(\xi) = \sum_{k=0}^{\infty} a_k\, \xi^{k-1} = \sum_{k=1}^{\infty} a_{k+1}\, \xi^k, \tag{2}$$

then

$$(1-\xi)\, \phi(\xi) = \sum_{k=0}^{\infty} a_{k+1}\, \xi^k - \sum_{k=1}^{\infty} a_k\, \xi^k = a_0 + \sum_{k=0}^{\infty} \Delta a_k\, \xi^k, \tag{3}$$

where, $\Delta a_k = a_{k+1} - a_k$, $k = 0, 1, 2, \ldots$ are finite differences of the first order of the coefficients $a_k$. Applying the Euler-Abel transformation to the power series $\sum_{k=0}^{\infty} \Delta a_k\, \xi^k$, **p** times, and after some manipulations we obtain

$$\sum_{k=0}^{\infty} a_k \xi^k = \sum_{i=0}^{\infty} \Delta^i a_0 \frac{\xi^i}{(1-\xi)^{i+1}} + \left(\frac{\xi}{1-\xi}\right)^p \sum_{k=0}^{\infty} \Delta^p a_k \xi^k, \tag{4}$$

where $\Delta^0 a_0 = a_0$. Equation (4) becomes meaningless when $\xi = 1$, so, by setting $\xi = -t$, we obtain the Euler-Abel transformed series as

$$\theta E_n(t) = \sum_{k=0}^{\infty} \Delta^i a_0 \frac{t^i}{(1-t)^{i+1}} + \left(\frac{t}{1-t}\right)^p \sum_{k=0}^{\infty} \Delta^p \left[(-1)^k a_k\right] t^k \tag{5}.$$

Returning to the earlier variable, $\xi$, we obtain

$$\theta E_n(\xi) = \sum_{i=0}^{p-1} (-1)^i \Delta^i a_0 \frac{\xi_i}{(1+\xi)^i} + \left(\frac{\xi}{1+\xi}\right)^p \sum_{k=0}^{\infty} (-1)^{k+p} \left[\Delta^p a_k\right] \xi_k, \tag{6}$$

where $\Delta^p a_k = \Delta^{p-1} a_{k+1} - \Delta^{p-1} a_k$.

Any order difference $\Delta^p a_k$, can be written as linear combination

$$\Delta^p a_k = \sum_{i=0}^{p} (-1)^{p-i} \binom{p}{i} a_{k+1}, \text{ where, } \binom{p}{i} = \frac{p!}{i!(p-i)!}.$$

The second step is to apply Pade' approximation to the Euler-Abel transformed series, Equation (6).

**Acknowledgments:** We thank the referee for his/her valuable comments which improved the paper.